\documentclass[twocolumn,preprintnumbers,amsmath,amssymb]{revtex4}

\usepackage{graphicx}
\usepackage{dcolumn}
\usepackage{bm}
\newcommand{\etal}{\emph{et al.}}

\begin{document}      
\title{A suggested 4$\times$4 structure in underdoped cuprate superconductors: a Wigner supersolid.} 
\author{P. W. Anderson}      
\affiliation{
Department of Physics, Princeton University, Princeton, N.J. 08544, U. S. A.
}

\date{\today}      

\begin{abstract}
A wave function is proposed for the ``$4\times 4$" inhomogeneous 
structures observed on cuprate superconductors.  It is based on the 
Gutzwiller-RVB technique proposed  in recent papers, and consists of 
a Wigner solid of hole pairs embedded in a sea of $d$-wave spin singlet 
pairs.  Arguments are given that the nodal quasiparticles may remain 
unscattered  and even superconducting on such a structure.
\end{abstract}
\maketitle                   
A number of recent STM experiments on underdoped cuprate
superconductors have shown evidence of  structure with a Bravais
lattice close to $4a\times 4a$ in the CuO$_2$ planes. A possibly similar 
structure has been
observed in vortex cores in BSSCO, with a fractionally larger
lattice constant; and we also call attention to the often-noticed
anomalies of $T_c$ and other parameters in LSCO at a doping of  $x$ = $\frac18$,
which to my knowledge have never been satisfactorily explained, and
may be caused by the same structure.  Both the tunnelling spectrum 
and the $\frac18$ structure seem often to indicate that there is 
superconductivity at low temperatures in materials with these 
structures.

I propose here a microscopic description of these phases that is not
inconsistent with their arising at low doping levels in an RVB
superconductor-Mott insulator system.  I am suggesting that they are
in essence a two-dimensional crystal of singlet $d$-wave pairs of 
holes, commensurate in many cases with the underlying lattice, and 
existing within a background of a
$d$-wave RVB of singlet pairs.  (Which latter is our model of the electronic
nature of the pseudogap state in these materials.)  Our model differs 
in essential ways from those of Chen \etal~\cite{chen} and Lee \etal~\cite{lee}, but has 
in common that we all propose a  crystal of holes.  The model of Lee 
\etal~explicitly rejects hole pairs in favor of individual holes, but 
seems to us to fail to explain why the structure appears when the 
material is still superconducting and still exhibits quasiparticle 
nodes--clearly the pairs are a reality, in spite of the Coulomb 
repulsion between the two holes.  It also loses the feature of 
explaining the ``$\frac18$" phenomenon and of why the doping level at which 
these observations occur is never as low as $\frac{1}{16}$.  The model of Chen 
\etal~is based on a proposed symmetry between antiferromagnetism and 
$d$-wave superconductivity which many consider problematic, and sees 
the crystal as a ``Wigner crystal", which to my mind implies an 
insulating state.  I should note that in a number of  numerical 
simulations on underdoped multi-leg ladders (White and Scalalpino~\cite{white}),
$4\times 4$ squares centered on a hole pair in a single plaquette
are observed as one of a number of inhomogeneous states.

In some recent papers~\cite{pwa1,pwa2} the author and collaborators have
returned to the early insight~\cite{zhang} that the ground state of the
CuO$_2$ planes in the cuprate superconductors can be modelled as a
Gutzwiller projection of a $d$-wave BCS superconducting state.  Our
point of view is to construct an effective Hamiltonian that operates
only within the manifold of ``lower Hubbard band" states -- in first
approximation, the ``$t$-$J$'' Hamiltonian -- and then to recognize that the
eigenstates of such a Hamiltonian must be general Gutzwiller
projected states.  We treat the states before projection by a
Hartree-Fock-BCS approximation, that is we find variationally the
best product of one-electron functions possible.  As in Hartree-Fock,
the mean field equations which arise from the variational procedure
also specify the quasiparticle excitations and their energies -- that
is, there is effectively a Koopman's theorem for this system.

In studying point-contact tunneling with this technique~\cite{pwa2} we found
it useful to use a formulation of the ground state wave-function
which to our knowledge was first given by Laughlin~\cite{laughlin}:
\begin{equation}
|\Psi \rangle = \exp(iS) \hat{P}_G\times[Z]^{n_{pairs}}\times |\Phi_{BCS}\rangle,
\label{Psi}
\end{equation}
where $n_{pairs}$ is the number of hole pairs.

Here, the Gutzwiller projector
\begin{equation}
\hat{P}_G = \frac12 \prod_{i}[1-n_{i,\uparrow} n_{i,\downarrow}]
\label{PG}
\end{equation}
is a projector that removes all doubly-occupied sites but otherwise
leaves amplitude and phase relations unaltered in the $d$-wave BCS wave
function $|\Phi_{BCS}\rangle$.  The product in Eq. \ref{PG} is over all sites. The BCS
function is that appropriate to the Fermi level for $1-x$ electrons,
where $x$ is the doping fraction.  $|\Psi\rangle$ is assumed to be determined
variationally as in Ref. \cite{pwa1}.  Finally, the canonical transformation
$\exp(-iS)$ transforms the true Hamiltonian into the projected form of
the $t$-$J$ Hamiltonian, and correspondingly its inverse transforms the projected
wave function into the true one.

Laughlin's innovation~\cite{laughlin} was to make explicit the ``fugacity factor" $Z$
raised to the number of hole pairs, which is necessary to adjust the
populations of the electron states on a given site.  It is easily
calculated that in a uniform state $Z = 2x/(1+x)$, with $x$ the doping fraction.
In Refs. \cite{pwa1,zhang,white} this population
adjustment is done by fiat.  In Ref. \cite{pwa2} the $Z$ factors were, further,
incorporated into the BCS function, which modifies the definition of
quasiparticle excitations in such a way as to give good agreement
with tunneling measurements, but we emphasize that the wave function
assumed there is just Eq. \ref{Psi}.  This variant of the wave function makes 
clear that $Z$ plays the role of a Bose condensate amplitude -- though we 
should emphasize that our theory does not involve assuming a ``holon 
condensate."

\begin{figure}[h]			
\includegraphics[width=7cm]{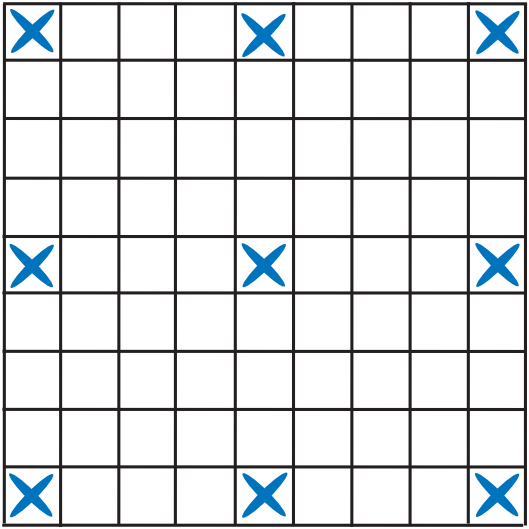}
\caption{\label{fig4by4}
Proposed 4$\times$4 structure.  Plaquets with crosses are nominal sites of hole pairs.
The structure may also be thought of as ``columnar'' valence bond structure commensurate with
hole-pair ``liquid crystal''.}
\end{figure}

What is being proposed here is that we approximate the wave function
for a density wave or Wigner-like crystal as described above not primarily by
changing the BCS function in Eq. \ref{Psi} but by requiring the fugacity $Z$ to
vary from site to site.  Its logarithm is the chemical potential for 
holes and thus must track the Madelung potential of the superlattice. 
The average fugacity is arranged to lead to the
correct doping level overall, but (for instance)  $Z$ may be taken to 
have one value, $Z_1$, on the central four sites of a 4$\times$4 square plaquette, and
another, smaller value $Z_2$ on the remaining 12 sites (or, if
desired, one may specify still a third value for the corners.)  Then
the hole density will be a maximum on the central plaquette, and will
be small in the intermediate regions.  The contrast in $Z$ values 
cannot in the end be very great, because $Z_{max}$ will likely not be 
greater than the value at optimal $T_c$ , about $\frac13$ (see Fig. 1).

The gap equations Eqs. \ref{Psi} and \ref{PG} which result from variation of  
$|\Psi\rangle$ are,  in the renormalized mean-field theory approximation of Ref. \cite{pwa1},
\begin{equation}
\Delta({\bf k}) = J(2-Z)^2 \sum_{\bf k'} \frac{\Delta({\bf k'})}{E({\bf k'})}
\label{Dk}
\end{equation}

\begin{equation}
\xi({\bf k}) = Z\epsilon({\bf k}) + J(2-Z)^2 \sum_{\bf k'} \gamma({\bf k}) \frac{\xi({\bf k'})}{E({\bf k'})},
\label{zeta}
\end{equation}
Here, $\gamma({\bf k}) = \cos(k_x - k_x') + \cos(k_y- k_y')$
and 
\begin{equation}
E({\bf k}) = \sqrt{\xi({\bf k})^2 + \Delta({\bf k})^2}
\label{Ek}
\end{equation}
is the standard BCS expression.  The important thing to note is that the
true kinetic energy $\epsilon(\bf k)$ is renormalized relatively by a
factor $Z/(2-Z)^2$ relative to $J$.  The underdoped regime is
defined by $Zt\ll 4J$, so that over most of the region near the Fermi 
surface, $E({\bf k})$ is relatively weakly dependent on $Z$.
Thus to a zeroth-order approximation the gap
equations Eq. \ref{Dk} are not affected by periodic spatial variation of $Z$,
justifying our basic \emph{ansatz} that the function before projection may
be assumed not much changed.  This signals the important fact that
the CDW lives in an RVB background, not in a conventional band or in
an antiferromagnetically (or even spin-glass) ordered state. The fact
that $\frac18^{\mathrm{th}}$ of a hole relative to the Mott-Hubbard insulator
corresponds to a unit cell of 16 sites does not add up in
conventional band theory.

Of course, the periodic variation of $Z$ will cause Bragg scattering of
the quasiparticles, which can be represented by dividing the spectrum
up into 16 sub-bands and describing the perturbation as a matrix in
band indices which opens Bragg-scattering gaps at the sub-zone 
boundaries.  I don't feel that Fermi surface nesting plays much of a 
role here.   Treating the resulting self-consistency problem 
accurately is beyond the patience or
ability of the present author to solve directly.  But there are a 
number of insights one can recognize.  First,  there will only be one of these subbands
which contains the nodal quasiparticles, and the lowest part of the
spectrum around the nodes will be relatively little scattered, since
the density of final states goes to zero as $E^2$ at low $E$.  The
experimental observation that the nodes survive and are coherent in 
the density-wave state is thus confirmed.  Second, we can expect the spectrum near the
antinodes to be severely broken up into rather flat ``optical" bands,
because both the gap and the underlying kinetic energy are rather
flat in this neighborhood.  The superfluid stiffness coming from
these portions of the zone will be severely reduced by this
scattering, and these excitations may move diffusively rather than
coherently.

Either from this insight, or on general principles, we can expect
that the superfluid stiffness $\rho_s$, which in the pure case is
known to be proportional to $Z$,  will be reduced more by the low
values of $Z$ between the peaks than by the higher value at the peaks,
and since $T_c$ is proportional to superfluid stiffness~\cite{pwa1}, $T_c$ will be
degraded.  The nodal quasiparticles, which are responsible for the
temperature dependence of $\rho_s$, are little changed.

It remains to discuss the motivation for this structure.  It is
obviously the long-range Coulomb interaction that furnishes the
energy gain, and the stiffness of the hole wave function which 
opposes the deformation.  We can suppose that $Z$  
is unlikely to exceed its value for optimal doping of 20$\%$, 
about 0.33, on the central plaquette, so that the 
decrease on the periphery is not severe -- I estimate 0.2 or so, or $x$ = 0.1.

The energy gain is something like the Madelung energy of the charge 
distribution of the ``liquid crystal"  which will be of the order of 
the square of the charge contrast divided by the lattice constant 
(and corrected by the dielectric constant).  Thus it scales as
\begin{equation}
E_{coulomb} \propto (\delta Z)^2\times 1/d
\label{Ec}
\end{equation}
where $d$ is the  superlattice constant.

The excess kinetic energy, on the other hand, will contain a 
gradient squared which is proportional to $(\delta Z/d)^2$.  Clearly 
there is some $d$ large enough so that the Coulomb energy gain wins. 
It is necessary that $d^2 = 2/x$.  Otherwise, the Madelung estimate 
is not correct, and we would have to include the Coulomb repulsion of the 
extra pairs.  Numerical estimates show that it is reasonable that the 
opposite variations of Coulomb and kinetic energy would match for 
$d = 4a$.

The above is not a complete theory.  In particular, we have not 
calculated the energetics explicitly.  But it is based on an explicit 
wave function, so that numerical calculations can be carried out when 
desired.  It does have the advantage that it leads to an explicitly 
superconducting state closely related to the uniform state which 
explains many of the properties of the superconductors.  There are 
several further questions which remain.

First, we need a reason why the lattice constant seems often to be 
slightly larger than 4.0, according to Fourier transforms of the 
tunnelling current.  The only real-space images of the microscopic 
structure (Davis~\cite{davis}) show however that the structure is made up of 
domains which are not continuously connected.  The packing of such a 
granular structure will not be perfect and will have a smaller 
average wave number.  We can speculate that commensurability with the 
lattice favors a definite hole density -- $\frac18$, we suppose -- and that 
the grain  boudaries adjust the net charge.  Thus the deviation from 
$4a$ provides suggestive evidence for, rather than against, the lattice 
of hole pairs.

A second puzzle can be resolved in much the same way. The vortex core 
state at optimal doping  exhibits a similar modulation, albeit with 
a lattice constant even larger than the previous one -- about 4.5$a$ is 
quoted.  In the vortex core the superfluid stiffness is reduced by 
the large supercurrent.  I expect the amplitude stiffness to 
gradients of $Z$ and the phase stiffness which causes supercurrents to 
be the same, and if one is reduced the other will be, and again the 
Coulomb energy will be favored relatively to the kinetic energy.  But 
here we expect the doping to be farther from $\frac18$ and the grains to be 
even smaller, hence the larger deviation from 4.0.

I would like to acknowledge the sharing of his data and his 
interpretations thereof by J. C. Davis, and useful discussions with N. P. 
Ong.

\end{document}